\documentclass[12pt]{article}

\usepackage{graphicx}
\usepackage{txfonts}
\usepackage{graphicx}
\usepackage[english]{babel}

\usepackage{indentfirst}
\usepackage[utf8]{inputenc}
\usepackage{natbib}
\usepackage{color}
\usepackage{hyperref}
\usepackage{subcaption}
\hypersetup{
    colorlinks=true,
    linkcolor=blue,
    filecolor=magenta,      
    urlcolor=cyan,citecolor=blue}

\begin{document} 
\pdfoutput=1
  
\title{Correlations between planetary transit timing variations,
transit duration variations and brightness fluctuations due to exomoons}
 \maketitle
 \centering
   ${K.E. Naydenkin}^1$ 
   and
      ${D.S. Kaparulin}^2$ \\
   ${}^1$ Academic Lyceum of Tomsk, High school of Physics and Math, Vavilova 8\\
rtut654@yandex.ru\\
${}^2$ Tomsk State University, Physics Faculty, Lenina av. 50\\ dsc@phys.tsu.ru
 
  \abstract
  
   {Modern theoretical estimates show that with the help of real equipment we are able to detect large satellites of exoplanets (about the size of the Ganymede), although, numerical attempts of direct exomoon detection were unsuccessful. Lots of methods for finding the satellites of exoplanets for various reasons have not yielded results. Some of the proposed methods are oriented on a more accurate telescopes than those exist today.
In this work, we propose a method based on relationship between the transit timing variation (TTV) and the brightness fluctuations (BF) before and after the planetary transit. With the help of a numerical simulation of the Earth-Moon system transits, we have constructed data on the BF and the TTV for any configuration of an individual transit. The experimental charts obtained indicate an easily detectable near linear dependence. Even with an artificial increase in orbital speed of the Moon and high-level errors in measurements the patterns suggest an accurate correlation.   
A significant advantage of the method is its convenience for multiple moons systems due to an additive properties of the TTV and the BF. Its also important that mean motion resonance (MMR) in system have no affect on ability of satellites detection.  
At the end we tried to retrieve parameters of the Moon (without additional satellites) on the basis of the simulated photometric curves. The results perform errors at the level of $\approx 0.45$ for the mass and $\approx 0.24$ for the radius with TTV measurements accuracy =60s and Kepler-like noise for photometry.

}
 
 \section{Introduction }

A bit more than fifteen years ago first constructive statements about satellites of extrasolar planets began to appear. Since then a large number of techniques aimed at exomoons search has been published. Based on time variations (Sartoretti \& Schneider \cite{sartoret} ; Lewis et al. \cite{lewis}; Kipping \cite{kipp1} ), on photometry analysis (Sartoretti \& Schneider \cite{sartoret}; Brown et al. \cite{brown}; Szabó et al. \cite{szab}; Charbonneau et al. \cite{charb}; Pont et al. \cite{pont}; Tusnski \& Valio \cite{tusnski}; Kipping \cite{kipp3}), on spectroscopic observations (Williams \& Knacke \cite{will}; Simon et al. \cite{simon1}; Robinson \cite{robin}; Zhuang et al. \cite{zhuang}; Heller \& Albrecht \cite{heller1}) and others.  Actually, none of these methods performed a reliable evidence of receiving a strong signal from satellites of exoplanets. Partially that happened because about a half of these research are keen on the future space observatories: PLATO telescope (Hippke \& Angerhausen \cite{hippk}) and CHEOPS (Simon et al. \cite{simon2}) and can't achieve necessary accuracy with the help of modern equipment. The same problems have spectroscopic studies based on Rossiter-McLaughlin effect. Some methods like rooted in strong computer treatment (Kipping et al. \cite{kipp5}) or in analysis of TTV-TDV diagrams (Heller et al. \cite{heller2}) may be used with current data from Kepler Space Observatory (or ground-based telescopes). At the same time the structure of TTV-TDV figures is very complex and it's difficult to build them with a small amount of planet-satellites transits. It's much easier to make a conclusion about the existence of exomoons around an exoplanet based on linear relationship.
In this paper we present a method where quasilinear connection was found between time transit variations and brightness fluctuations near the planetary transits. This technique is applicable to the Kepler data and gives an opportunity to detect groups of satellites as well as planet-one satellite systems.

The actuality of the study was highlighted by the fact that habitable zones are not the only places where life can be found. Tidal forces of a gas giant and close satellites can organize suitable life conditions on gas giant's moons. Examples of such behaviour can be found in our own Solar system. The Jupiter's satellite Europa and the Saturn's Enceladus may have microbial lifeforms under ices of their surface.

So, at the present moment of time we still stand near the front door of the revolution in the sphere of habitable satellites studying. The discoveries of satellites outside our home expand our possibilities in search for life.
 
 \section{TTV and brightness fluctuations due to an exomoon }

Let's assume a faraway spectator observes a planetary system of a certain star. So his visual ray is located accurate in the plane of the system. Our spectator could make ephemerides for transits of one chosen planet. In the case, the time of begining and ending of every new transit will shift comparing with the ephemerides. The difference between observed and predicted times of a transit is known as transit timing variation or TTV (Sartoretti \& Schneider \cite{sartoret}) . The variation of the overall transit time is known as time duration variation or TDV (Kipping \cite{kipp1}).

Actually, such variations of the planet may be caused by variety of things. First of all, we can observe a perturbing planet with an unstable orbit. Different effects were studied as an effective sources of timing variations:  
an effect of parallax (Scharf \cite{schar}), trojan bodies (Ford and Holman \cite{ford}), stellar proper motion (Rafikov \cite{rafik}), Kozai mechanism induced eccentricity variation (Kipping \cite{kipp0}), apsidal precession (Jord\'an and Bakos \cite{jord}, stellar spots and darkening (Alonso et al., \cite{alonso}), planetary in-fall (Hellier et al. \cite{hell}), the Applegate effect (Watson and Marsh \cite{watson}) and stellar binarity (Montalto \cite{monta}).

On the other hand, one of the reasons of such behaviour would be the existence of exomoons around this planet. Due to this fact, TTV and TDV occur. The structure of the transit is presented in (Figure 1) .

 \begin{figure}[h]
\centering
\includegraphics[width=0.40\textwidth]{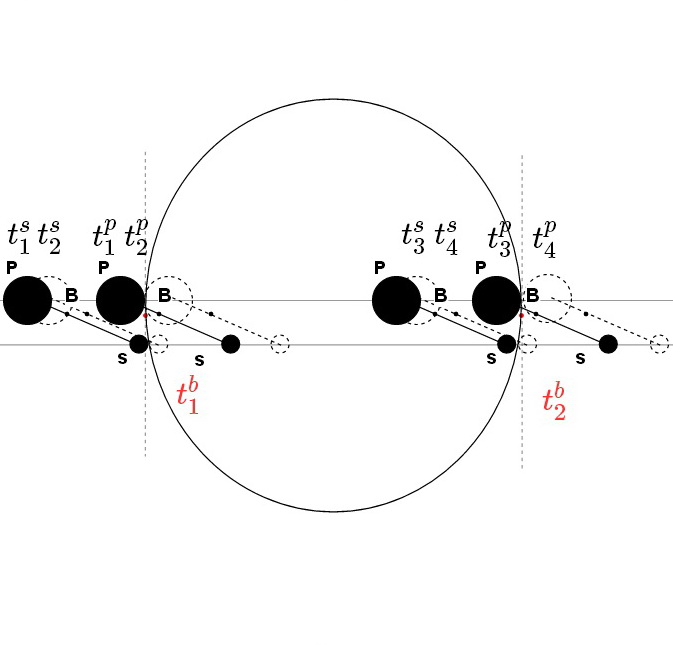}
\caption{  An illustration of the transit of a planet with one satellite }
\label{fig:figure2}
\end{figure}

In Figure 1, we illustrate the main stages of the transit of an exoplanet system with one satellite. Here, the big white circle denotes the star, the middle circle denotes the planet (p) and small circle denotes the satellite (s). The barycenter (b) of the planet - satellite system is shown as the small dot on the line
connecting the planet and satellite. The times $t^s_1,t^b_1,t^p_1 (t^s_2,t^b_2,t^p_2)$ correspond to the moments when the center of the satellite, barycenter and the center of the planet enter (come off) the stellar disc.

The satellite influences the transit in two ways: (i) it changes the moments of transit's beginning and end (causes transit timing variation), and (ii) decreases the star's flux beyond the planetary transit. The transit timing variations of begin and end of transit, $\delta t^p_1$ and $\delta t^p_2$, are defined by
\begin{equation}\label{ttvs}
\delta t^p_1=t^p_1-t^b_1\,,\qquad \delta t^p_2=t^p_2-t^b_2\,.
\end{equation}
The difference $ t_{\mathrm{tdv}}=\delta t^p_2-\delta t^p_1$ is known as time duration variation. Transit duration variation is caused by the relative motion of the planet and the satellite during transit. 
It's known that for orbital stability (stable orbit in Hill's sphere of the planet) satellite's period $p_s$ and $p_p$ - orbital period of the planet around the host star should satisfy the inequality: $p_p\leq9p_s$ (Kipping \cite{kipp1}). On the other hand, for the planet in the habitable zone orbiting the mid-main sequence star the transit duration is two or three orders less than the orbital period of the planet (Kipping \cite{kipp1}). In our work, we assume that
\begin{equation}\label{smm}
t^p_2-t^p_1\ll p_s \ll p_p\,.
\end{equation}
Upon this assumption, the orbital speed of the planet $v_p$ and the orbital speed of the satellite $v_{sp}$ can be expressed as
\begin{equation}\label{smm}
v_p=\frac{r_c}{t^p_2-t^p_1}\,,\qquad v_{sp}=-\frac{m_sv_p}{m_p+m_s}\frac{t_{tdv}}{t_{ttv}}.
\end{equation}

Here, $r_c$ is the star's radius and $m_p,m_s$ are the masses of the planet and satellite. If the relative potion of the planet and satellite does not qualitatively change during transit ("frozen" moon), then $t_{tdv}=v_{sp}=0\,.$

By the time Kepler Space Observatory was launched it was clear that the accuracy of a single frame would be worse than a brightness decreased by Moon in the center of Solar disc. Additionally, taking into account errors in measurements and star limb darkening it's almost impossible to detect satellite size of Ganymede around an exoplanet with only single frames.  That's why more complicated and sensitive methods analyze relationship between different quantities like $\mathrm{TTV}$, $\mathrm{TDV}$ and photometric data.
In the present work, we propose to make a conclusion about the presence of an exomoon on the basis of correlation between the transit time variation and variations of star's flux beyond the planetary transit. As the photometric characteristics of the transit we chose average star's flux withing the time intervals with duration $t_{int}$ before and after the planetary transit:

 \begin{equation}\label{J1J2}
F_1=\frac{1}{t_{int}}\int_{t^p_1-t_{int}}^{t^p_1} F(t)dt \,,\qquad F_2=\frac{1}{t_{int}}\int_{t^p_2}^{t^p_2+t_{int}} F(t)dt \,.
\end{equation}
Here, $F(t)$ is star's flux at the moment $t$; The total star flux is normalized to unity, so beyond the transits \mbox{$F(t)=1$}.

As a main observation we would like to reveal that the transit timing variation and brightness fluctuations are correlated. One can get that
\begin{equation}\label{dJ}
\delta F\equiv F_2-F_1=\frac{1}{t_{int}}\left(\int_{t^p_2}^{t^p_2+t_{int}}  F(t)dt-
\int_{t^p_1-t_{int}}^{t^p_1}  F(t)dt\right)\,,
\end{equation}
where $\delta F(t)$ is the flux variation at the moment $t$. Assuming that both the exomoon and star have a spherical form, and radius of the exomoon is much smaller that of the star, $r_s\ll r_c$, we can estimate
 
 \begin{equation}\label{dJt}
\delta F(t)=\frac{r^2_s}{r^2_c}I(r_d(t))\,,
\end{equation}
where $I(r)$ is the surface brightness of the star, and $r_d$ is a relative distance between the visible position of the satellite and the center of the star's disc. Suppose now that the satellite crosses the star's disc with the constant speed $v_p$. Then substituting (\ref{dJt}) to (\ref{dJ}) and switching to integration by $r_d$, we get
\begin{equation}\label{dF}
\delta F=\frac{r^2_s}{r_c v_s t_{int}}\int_{r^0_d}^{r^1_d}I(r_d)dr_d=\alpha\int_{r^0_d}^{r^1_d}I(r_d)dr_d\,.
\end{equation}
 Here, the initial and final positions of the satellite, $r_0^d$ and $r^1_d$, are given by
\begin{equation}\label{r01}\begin{array}{ll}
r_d^0=\left\{%
\begin{array}{ll}
    \displaystyle 1+\beta t_{{ttv}}\,, & t_{{ttv}}>0\,;  \\[1mm]
    \displaystyle {min}\{1,1-\beta t_{{ttv}}+t_{int} \frac{v_s}{r_c}\}\,,& t_{{ttv}}<0\,; \\
\end{array}%
\right.\\[7mm]
r_d^1=\left\{%
\begin{array}{ll}
    \displaystyle {min}\{1,1+\beta t_{{ttv}}+t_{int} \frac{v_s}{r_c}\}\,, & t_{ttv}>0\, \\[1mm]
    \displaystyle 1-\beta t_{ttv}\,, &   t_{ttv}<0\,,
\end{array}%
\right.
\end{array}\end{equation}
Here, $m_p$ and $m_s$ are being the masses of the planet and the satellite, and the notation is used
\begin{equation}\label{beta}
\beta=\frac{v_s}{r_c}\frac{m_s+m_p}{m_s}\,.
\end{equation}
 
 If the relative speed of a host planet and exomoon $v_{sp}=v_s-v_p$ is much less than the orbital speed of the planet $v_p$, with account (\ref{smm}) we can suggest the following approximation for $\beta$:
\begin{equation}\label{beta1}
\beta=\frac{v_p+v_{sp}}{r_c}\frac{m_s+m_p}{m_s}\approx \frac{v_p}{r_c}\Big(\frac{m_s+m_p}{m_s}-2\frac{v_p}{r_c}t_{{tdv}}\Big)\,.
\end{equation}
In the last formula, all the values except the ration $m_s/m_p$ can be found either from photometric measurements or the orbital parameters of the planet. The latter fact suggests that there is functional dependence between brightness fluctuations (BF), and transit timing variations $t_{{ttv}},   t_{{tdv}}$. This dependence has the form
\begin{equation}\label{dFab}
\delta F\equiv Y(\alpha,\beta,v_p, t_{{ttv}}, t_{{tdv}})=\frac{\alpha}{v_p t_{int}}\left(J(r^1_d)-
J(r^0_d)\right)\,,
\end{equation}
where $J(r)=\int I(r)dr$ is the antiderivative of $I(r)$, $r^1_d$ and $r^0_d$ are defined in (\ref{r01}); the constants $\alpha$ and $\beta$ are defined in (\ref{dF}) and (\ref{beta1}).
 
 The function $Y$ can be found for any reasonable surface brightness distribution of the star. For example, if surface brightness distribution of the star has the form
\begin{equation}\label{ic0c2c4}
I(r)=c_0 + c_2\sqrt{(1-r^2)}\,,
\end{equation}
with $c_0$ and $c_2$ being the constant coefficients,
the function (\ref{dFab}) reads
\begin{equation}\label{dFab1}\begin{array}{ll}
\delta F=&\displaystyle \frac{\alpha}{2 v_p t_{int}}\Big(c_0x+\frac{c_2}{2}\arcsin(x)+\frac{c_2}{2}\sqrt{1-x^2}\Big)\Big|_{r^0_d}^{r^1_d}\\[3mm] 
\end{array}\end{equation}
Figure 2 shows a typical plot of this function if $t_{{int}}\leq t_{{crit}}$. Figure 3 shows a typical plot of this function if $t_{{int}}>t_{{crit}}$. Here, the critical time $t_{{crit}}\approx r_c/v_p$ is longest time the satellite can stay on project star disc before or after the planetary transit. The blue curve 'fast moon' accounts the $t_{{tdv}}$-effect caused by the motion of the moon. The two branches of the function correspond to positive and negative signs of $t_{{tdv}}$ for a certain $t_{{tdv}}$. To make the difference between two branches of the graph more visible the orbital velocity of the moon was increased 20 times (or orbital period set to 1.37 days), for the real plot two branches merge to each other. The red curve 'frozen moon' corresponds to the case when the relative position of the satellite and the planet does not change, or equivalently, $t_{{tdv}}=0$. We will mostly be concentrated on the case of frozen moon, because it is much simpler.

 The TTV-BF analysis applies to the planetary systems with multiple moons. The crucial observation in this case is that the total contribution to the BF and TTV is the sum of contributions of separate moons: 
\begin{equation}\label{dFdt}\begin{array}{l}\displaystyle
\delta F=(\delta F)_1+(\delta F)_2+(\delta F)_3+\ldots+(\delta F)_n\,,\\[3mm]
\displaystyle t_{{ttv}}=(t_{{ttv}})_1+(t_{{ttv}})_2+(t_{{ttv}})_3+\ldots+(t_{{ttv}})_n\,,
\end{array}\end{equation}
where $n$ is the number of moons. Due to additive properties of TTV and BF, the total TTV-BF effect for the planetary system can be higher than from a single moon. We perform numerical simulations TTV-BF plots for some particular system with two moons.  

\section{Numerical simulations for TTV and BF}

In this section we present the results of numerical simulations for transit timing variations and brightness fluctuations. For simplicity reasons we assume that orbits of the satellite and the planet are circular and coplanar, and the transits are central. The bodies of the star, the planet and the satellite are supposed to be spherical. The motion of the planet and satellite(s) was computed in the approximation where the barycenter of the planetary system follows the Keplerian orbit around a host star and planetary satellites rotate around the barycenter. The radius of the host star is set $r_c=70000\,{km}$, the radius of the planetary orbit and orbital period are $a_p=150000000\,{km}$, $P_p=365.25\,{days}$. The surface brightness distribution has the form (\ref{ic0c2c4}) with $c_0=0.142$, $c_2=0.264$, and $c_4=0$. Thus, our model describes a Earth-like planet rotating a Sun-like star.

For the first simulation we took the Earth-Moon system, but with inclination of the Moon's orbit $i_s=0$ and zero eccentricity. Figures 2 and 3 perform results of this numerical modeling with a different integration times. The red curve in these figures corresponds to the "frozen" Moon model and the blue curve corresponds to the "moving" Moon. For clarity (to better visualize the expansion of the blue curves) we increased the orbital speed of the Moon up to 20 km per second. Figure 2 corresponds to $t_{int}$=6500 sec and Figure 3 corresponds to the critical integration time, $t_{crit} \approx t_{int} =$ 13000 sec. 
\begin{figure}[h]
\centering
\includegraphics[width=0.52\textwidth]{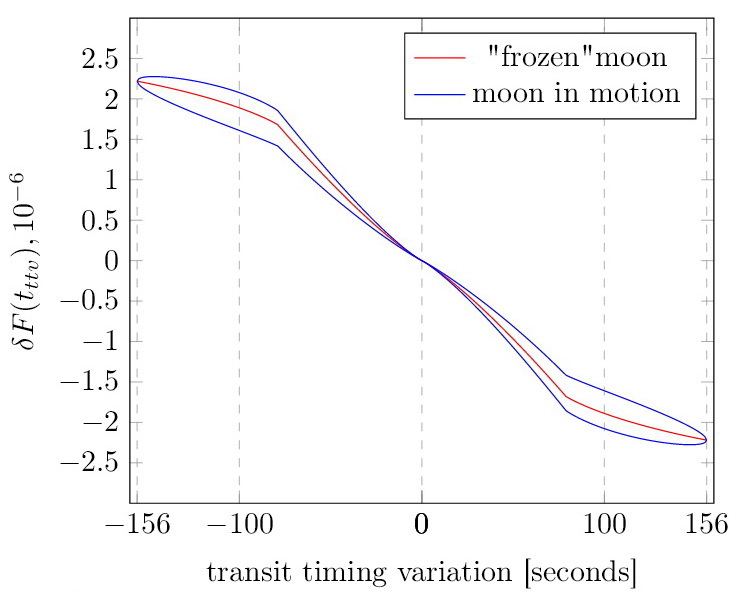}
\caption{Results of numerical modeling with Earth-Moon system parameters, $t_{int}$=6500 sec}
\centering
\label{fig:2a}
\end{figure}

\begin{figure}[h]
\centering
\includegraphics[width=0.48\textwidth]{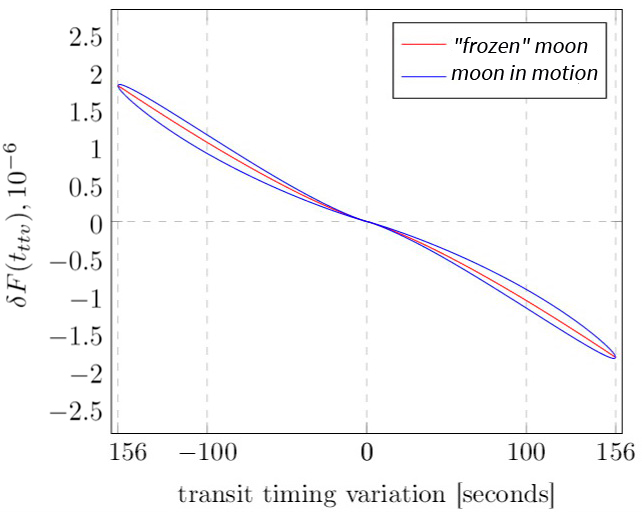}
\caption{Results of numerical modeling with Earth-Moon system parameters, $t_{int}$=13000 sec}
\centering
\label{fig:2b}
\end{figure}

So, when $t_{int}$ less critical , the diagram change its form (figure 2). Such a rapid curve shift occurs due to limb darkening function. For example, for $TTV>0$ some decrease in integration time (in this case $t_{int}<$13000 sec) leads to the fact that the satellite does not reach the closer and brighter area on the star disc. As a consequence the Moon makes much less contribution to $\delta F(t_{ttv})$.

The TTV-BF diagrams have a particular interest for the systems with multiple moons, where TTV-TDV dependence have complicated structure. Here, two cases seem to be different:

1 Multi-moons systems without orbital mean motion resonance (MMR).

2 Multiple moons in MMR.

In the first case (the system without MMR), its makes sense to speak only of the special zone, which limits the points chart. For example, (Figure 4a) shows a diagram of Earth-Moon system with the introduction of one more closer (than the Moon) companion ($a =200000 km, r_s =1450 km, m_s=1/95 m_p$). The blue line restricts the area obtained by plotting $\delta F(t_{ttv})$ for this system. Red dots obtained by numerical simulation of the transits follow the "frozen moons" model. On (Figure 4b )  satellites are transiting the Sun with their own speeds according to their orbits parameters. Figure 5a and 5b show the same retrieval but with a more distant (than the Moon) satellite, matches to Jupiter's Io ($a_s =420000 km, r_s=1820 km, m_s=1/67 m_p$).

In the second case (taking into account MMR), the diagrams represent curves instead of special zones (Figure 6,7). On the figure 6, the Moon was introduced in 2:1 MMR with a closer satellite. (Figure 6a) and (Figure 6a) are differ because of phase shift, in (Figure 6a) the satellite and the Moon always shifted by $\pi/2$ in the central moment of transit. 

So, the introduction of additional satellites virtually had no impact on the type of chart. In the first case (without MMR), contact zones are very similar to charts we made for a single moon.
Similarly, in the case of MMR charts practically do not change their appearance. That is, the proposed method can also be applied to systems of two satellites.
%
%______________________________________________________________

\section{Application to detection of exomoons}
In the second stage of the simulation we made an analysis of $\delta F(t_{ttv})$ diagrams. As it was shown in theoretical section (n. 2), brightness fluctuations function (\ref{dFab}) depends only on the two unidentified characteristics: the mass of the satellite ($m_s$) and its radius ($r_s$), which are contained in the parameters $\alpha $ and $\beta$. The third parameter, the semi-major axis of satellite's orbit, can not be found from the $\delta F(t_{ttv})$ charts. However, having a sufficient amount of transits we can estimate orbital radius on the basis of the maximum transit timing variation.This implies that 
$v_p t_{ttv} \frac{m_p+m_s}{m_s}\rightarrow a_s$ when $t_{ttv} \rightarrow max$. Thus, with the help of iterative approximations of experimental points on the diagram we can get the desired parameters of the satellite. At the same time, it is evident that the satellite orbit's radius will be found based solely on the maximum absolute amount of $t_{ttv}$.
Table 1 and 2 show the parameters of data retrieval in Earth-Moon system (without an additional satellite).

\begin{table}{Table 1.Retrieval of the parameters with iterative approximations }
\centering
 \begin{tabular}{c c c c} 
  \noalign{\smallskip}
 \hline
 \noalign{\smallskip}
Parameter & True  & Fitted \\ [0.5ex] 
 \hline
$  r_s , km $  & $1740 $    &  $ 1751  \pm 0.84 $ \\
$ m_p/m_s $ & $  81.3 $  & $80.4 \pm 0.075$ \\ [1ex] 
 \hline
 \end{tabular}
\end{table}

\begin{table}{Table 2. Parameters retrieval with $\delta t_{ttv}=60 s, \delta F=0.00006$}
\centering
 \begin{tabular}{c c c c c} 
  \noalign{\smallskip}
 \hline
 \noalign{\smallskip}
Parameter & True & Fitted \\ [0.5ex] 
 \hline
$ r_s , km$  & 1740  & $1302  \pm 310.8$ \\
$ m_p/m_s$ &   $81.3 $ & $ 36.4 \pm 12.2  $ \\  [1ex] 
 \hline
 \end{tabular}
\end{table}

%
%______________________________________________________________

\section{Conclusions}

 Based on the photometric curves of a limited number of Earth's transits (corresponding to the actual situation with the Kepler Observatory data), we tried to restore the real parameters of the Moon. As it was shown, the Moon's radius can be found with the accuracy for the radius about 590 km and for the mass $\approx 140  m_p/m_s$. It's important that in the case of linear brightness distribution $\delta F(t_{ttv}) \sim \alpha \beta$. So, the parameter $ {r_s}^2 m_s$ which equivalent to $\alpha \beta$ can be found with the best accuracy. Because of the fact that a real brightness distribution is very close to linear, result of $\alpha \beta$ parameter retrieval is close to the real value.
 Large errors in determining the mass associated with errors in measuring the position of the planet. The measurement errors depend on the stellar magnitude of the observed star. In this case, the accuracy used (1 min) can be achieved with ground-based telescopes. For example, during 2012 at Gemini North a 1 minute brightness integration was used for stars of 11-12 Kepler magnitude  (Rowe et al. \cite{Rowe}). 
 The accuracy taken for photometry ($\delta F=0.00006$) is close to the maximum for the Hubble. According to HST's official data (http://hubblesite.org/) ,
  photometry of the transiting planet HD209458b has achieved near-Poisson-limited precision (but in ten min. integration instead of one) with the largest non-random errors resulting from rapidly changing environmental conditions related to HST's low-Earth orbit.

The main advantage of the method is its accuracy increases with increasing of the number of satellites around the planet (because TTV and BF are additive values).
With the introduction of an additional moon at a distance greater or smaller than the radius of the Moon's orbit, the shape of the curve changes insignificantly and the accuracy of satellite detection multiplies. Even with an artificial increase in the speeds of the satellites (to the velocities close to the speed of the planet), the shape of the curve remains the same.
The points corresponding to the transits of the Earth with several satellites without resonance (MMR) are limited by the two-dimensional figure of "possible hit." In the case of satellite resonances, the graph has the form of a curve. The type of resonance determines the shape of the curve. For example, in the case of 2:1 MMR the function curve is enclosed. In the 3:1 case, the function has a form of an open curve.

 Thus, the proposed method with the use of the modern data can not achieve necessary accuracy in determining the parameters of a planetary system with a single moon. At the same time, regardless the number of satellites in system , the method makes it possible to find out the signs of their presence with high accuracy.

 \textit{Acknowledgments} 
 
 We would like to thank R. Heller for the helpful discussions. 
 Also we thank T. Yu. Galushina from Astronomy department of Tomsk State University for the useful comments on stellar mechanics.
 The work was partially supported by the Tomsk State University Competitiveness Improvement Program.

\begin{figure}[h]

\begin{subfigure}{0.5\textwidth}
\includegraphics[width=0.9\linewidth, height=5cm]{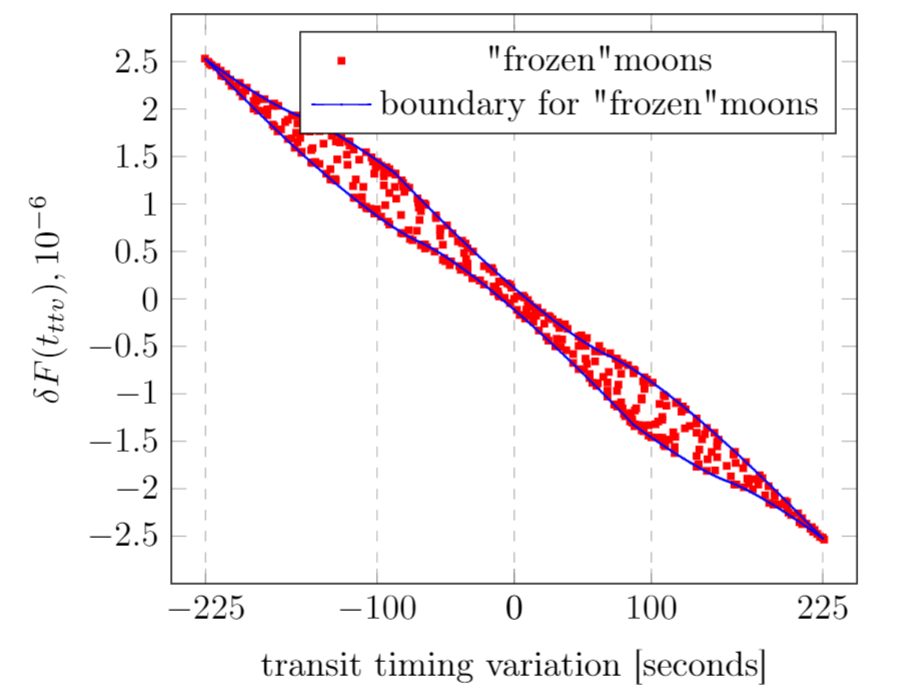} 
\caption{"frozen" moons model}
\label{fig:3}
\end{subfigure}
\begin{subfigure}{0.5\textwidth}
\includegraphics[width=0.9\linewidth, height=5cm]{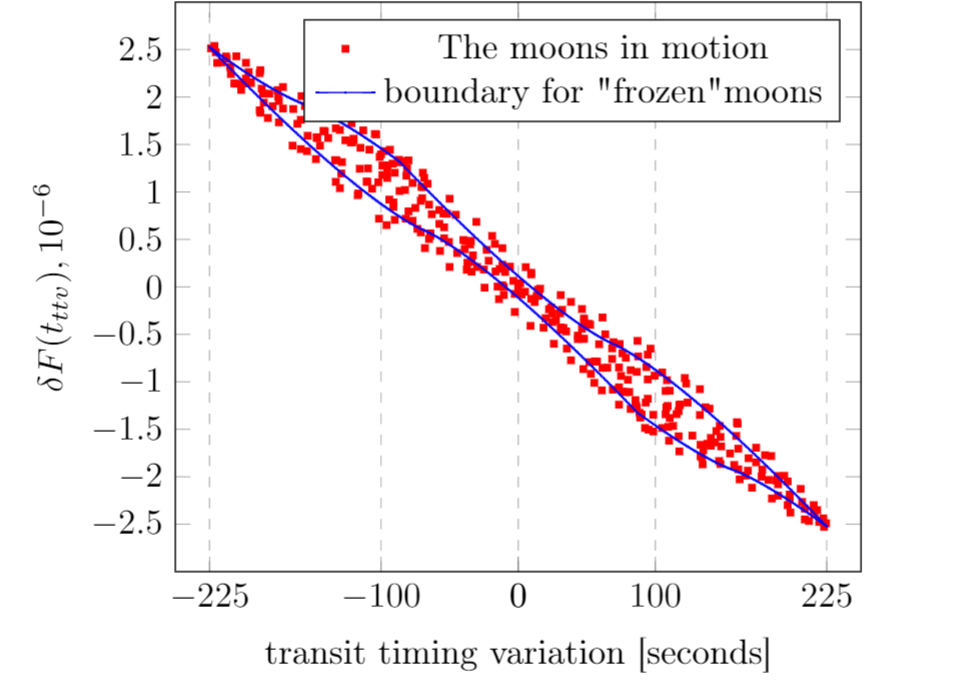}
\caption{normal orbital speeds}
\label{fig:subim2}
\end{subfigure}
\caption{Close companion
($a =200000 km, r_s =1450 km, m_s=1/95 m_p$)}
\label{fig:image2}
\end{figure}

\begin{figure}[h]
\begin{subfigure}{0.5\textwidth}
\includegraphics[width=0.9\linewidth, height=5cm]{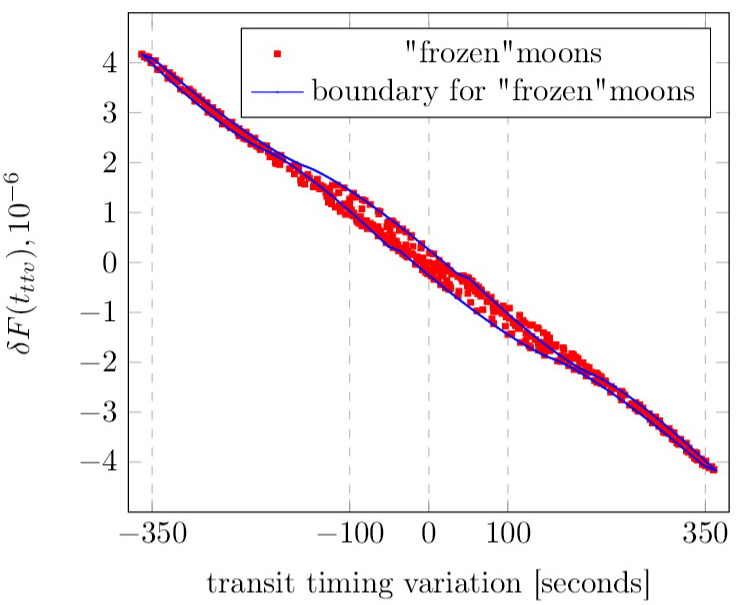} 
\caption{"frozen" moons model}
\label{fig:subim1}
\end{subfigure}
\begin{subfigure}{0.5\textwidth}
\includegraphics[width=0.9\linewidth, height=5cm]{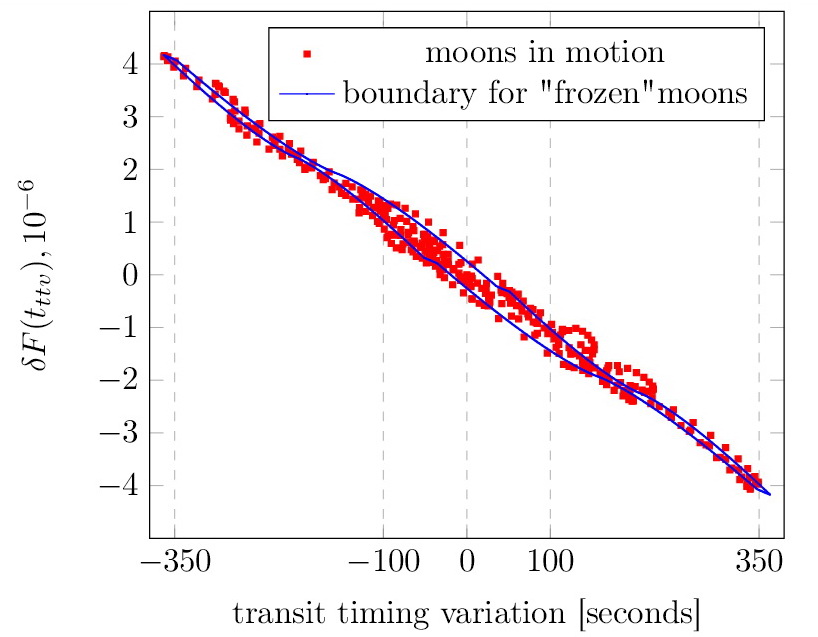}
\caption{normal orbital speeds}
\label{fig:subim2}
\end{subfigure}
\caption{The Moon with the Io-like additional satellite}
\label{fig:image2}
\end{figure}

\begin{figure}[h]
 \begin{subfigure}{0.5\textwidth}
\includegraphics[width=0.9\linewidth, height=5cm]{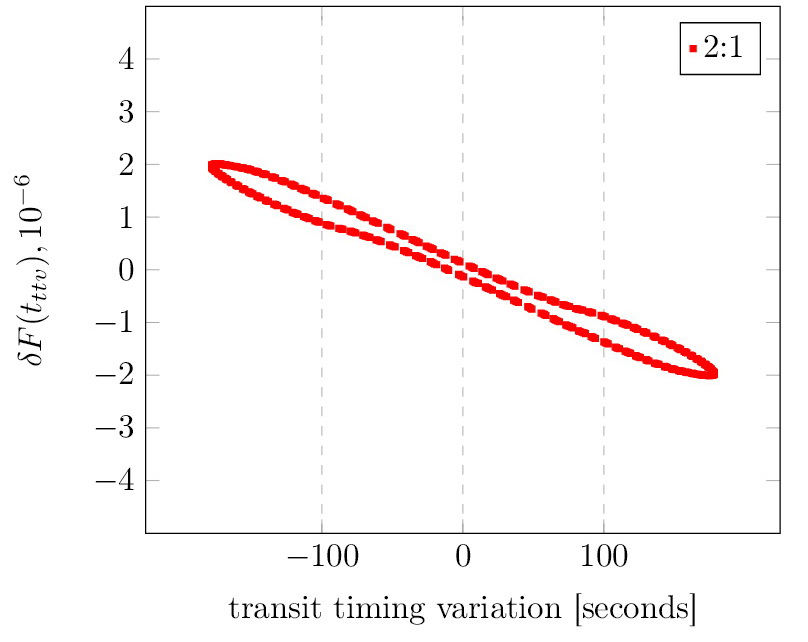} 
\caption{The Moon with a close companion in 2:1 MMR , initially shifted  by $\pi/2$}
\label{fig:subim1}
\end{subfigure} 
\begin{subfigure}{0.5\textwidth}
\includegraphics[width=0.9\linewidth, height=5cm]{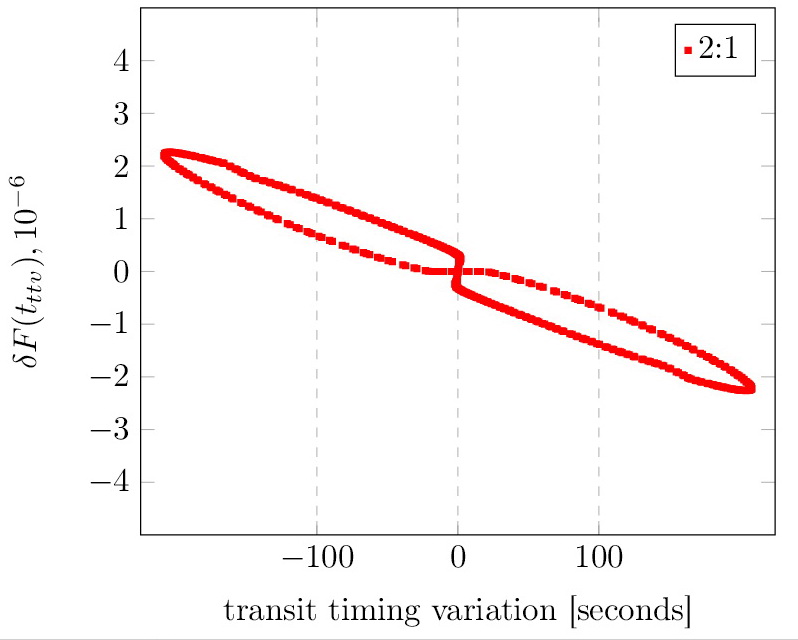} 
\caption{The Moon with a close companion in 2:1 MMR}
\label{fig:subim1}
\end{subfigure}
\caption{Close companion
($r_s =1450 km, m_s=1/95 m_p$) in 2:1 MMR}
\end{figure}

\begin{figure}[h]
\begin{subfigure}{0.5\textwidth}
\includegraphics[width=0.9\linewidth, height=5cm]{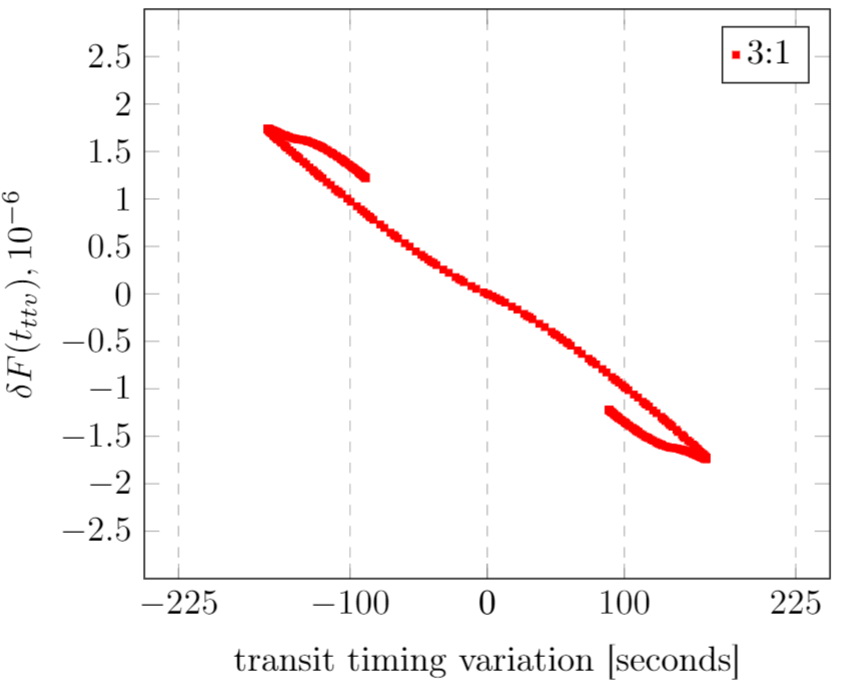}
\caption{The Moon with a close companion in 3:1 MMR}
\label{fig:subim2}
\end{subfigure}
\label{fig:image2}
\caption{Close companion
($r_s =1450 km, m_s=1/95 m_p$) in 3:1 MMR}
\end{figure}

\end{document}